\documentstyle[12pt]{article}
\newcommand{\beq}{\begin{equation}}
\newcommand{\eeq}{\end{equation}}
\newcommand{\bea}{\vspace{0.25cm}\begin{eqnarray}}
\newcommand{\eea}{\end{eqnarray}}

\newcommand{\r}{\mbox{{\boldmath
$\rho$}}}

\newcommand{\qb}{\mbox{{\bf
q}}}
\newcommand{\pb}{\mbox{{\bf
p}}}

\def\lsim{\mathrel{\rlap{\lower4pt\hbox{\hskip1pt$\sim$}}
    \raise1pt\hbox{$<$}}}         
\def\gsim{\mathrel{\rlap{\lower4pt\hbox{\hskip1pt$\sim$}}
    \raise1pt\hbox{$>$}}}         
\begin{document}
\thispagestyle{empty}
\vspace*{-2cm}
\begin{flushright}
{\bf\large
FZJ-IKP(Th)-2000-31\\}
\end{flushright}
 
\bigskip

\begin{center}

  {\large\bf
ON THE ENERGY LOSS OF HIGH ENERGY QUARKS  
IN A FINITE-SIZE QUARK-GLUON PLASMA
\\
\vspace{1.5cm}
  }
\medskip
  {\large
  B.G. Zakharov
  \bigskip
  \\
  }
{\it  Institut  f\"ur Kernphysik,
        Forschungszentrum J\"ulich,\\
        D-52425 J\"ulich, Germany\medskip\\
 L.D. Landau Institute for Theoretical Physics,
        GSP-1, 117940,\\ Kosygina Str. 2, 117334 Moscow, Russia
\vspace{2.7cm}\\}

  {\bf
  Abstract}
\end{center}
{
\baselineskip=9pt
We study within the light-cone path integral approach the induced gluon 
emission from a fast quark passing through a finite-size QCD plasma. 
We show that the leading log approximation used in previous studies 
fails when the gluon formation length becomes of the order of 
the length of the medium traversed by the quark.
Calculation of the energy loss beyond the leading log
approximation gives the energy loss which grows logarithmically with quark
energy contrary to the energy independent prediction of the leading
log approximation.
}
\pagebreak
\newpage

In resent years much attention has been attracted to the problem of
the induced gluon radiation from fast partons in a hot QCD medium 
(for a review, see \cite{BSZ}).
It is of great interest in connection with the current 
experiments at  SPS, RHIC, and future experiments at  
LHC on $A+A$ collisions since jet quenching due to the parton energy
loss can be a good probe of formation of a hot quark-gluon plasma (QGP).

Evaluation of the gluon emission from a fast parton in a medium requires
the understanding of the non-abelian analogue of the 
Landau-Pomeranchuk-Migdal (LPM) effect \cite{LP,Migdal}.
There are two approaches to the LPM effect in QCD:
the so-called BDMS approach \cite{BDMS} 
(see also \cite{BSZ,BDMPS}) based on the Feynman diagrammatic formalism, and 
the light-cone path integral (LCPI) approach developed in our paper 
\cite{Z1} (see also \cite{Z_YAF,Z2,Z3,Z4}). The BDMS approach neglects
the mass effects, and applies for large suppression of the radiation 
rate as compared to the Bethe-Heitler one. The LCPI approach applies for 
arbitrary strength of suppression.
For large suppression these approaches are equivalent \cite{BDMS,BSZ,W1}.
The probability of gluon emission in the BDMS and LCPI 
approaches is expressed through the solution of a 
two-dimensional Schr\"odinger equation with an imaginary potential.
This equation describes evolution of the color singlet $\bar{q}gq$ 
system in the medium. The potential is proportional to the cross section
for scattering of the $\bar{q}gq$ system on a medium constituent.
For the QGP the constituents  can be modeled as 
Debye-screened colored Coulomb scattering centers \cite{GW}.

In \cite{BDMS} the quark energy loss, $\Delta E$, has been evaluated 
analytically treating interaction of the $\bar{q}gq$ 
system with the Debye-screened centers in the Leading Log 
Approximation (LLA) which is equivalent to 
the harmonic oscillator approximation for the Hamiltonian of the $\bar{q}gq$
system. For a quark produced inside a finite-size QGP the BDMS prediction is
\beq
\Delta E_{BDMS}=\frac{C_{F}\alpha_{s}}{4}\frac{L^{2}\mu^{2}}{\lambda_{g}}
\tilde{v}\,,
\label{eq:1}
\eeq
where $L$ is the length of QGP traversed by the quark, $\mu$ is the 
Debye screening mass, $\lambda_{g}$ is the mean free path of
the gluon in QGP, $C_{F}$ is the color Casimir for the quark,
and the factor $\tilde{v}$ grows smoothly with $L$, at $L\gg\lambda_{g}$ 
$\tilde{v}\approx\log(L/\lambda_{g})$.
 
The energy independent $\Delta E$ (\ref{eq:1})
differs from that obtained recently by
Gyulassy, Levai, and Vitev \cite{GLV}. Calculating the Feynman diagrams 
for the single scattering (the first order $(N=1)$ in opacity) they have 
obtained 
\beq
\Delta E_{GLV}=\frac{C_{F}\alpha_{s}}{4}\frac{L^{2}\mu^{2}}{\lambda_{g}}
\log\frac{E}{\mu}\,.
\label{eq:2}
\eeq
Since the $\Delta E_{BDMS}$ should include the $N=1$ contribution
the contradiction between (\ref{eq:1}) and (\ref{eq:2}) 
at $E\rightarrow \infty$ seems to be 
surprising\footnote{Strictly speaking, the derivation of the BDMS formalism 
given in Ref. \cite{BDMS} is valid only when the number of rescatterings
is large.  However, since the formulas obtained are equivalent to 
those of the LCPI \cite{Z1} approach which is free from this restriction, 
it is clear that the BDMS prediction should contain the $N=1$ term.}.
By now there has not been given any explanation of this fact,
except the argument of the authors of Ref. \cite{GLV} that 
it can be connected with the neglect of the finite kinematic bounds
in the analysis \cite{BDMS}. However, it is clear that it 
cannot be important at $E\rightarrow \infty$.   

In the present paper we resolve the above puzzle of the discrepancy 
between the BDMS and GLV predictions.
We demonstrate that the absence of the logarithmic energy dependence
in (\ref{eq:1}) is connected with the fact that the LLA fails
when the gluon formation length becomes of the order of $L$. 
In this case the spectrum is dominated by the $N=1$ scattering 
which simply vanishes in the LLA. 
We show that if one uses the actual imaginary potential  
the energy loss grows logarithmically with quark energy. However,
the denominator in the argument of the logarithm is not the Debye 
mass as it is in (\ref{eq:2}).


We will work in the LCPI formalism \cite{Z1}.
The probability distribution of the induced gluon 
emission from a quark produced at $z=0$ can be written as \cite{Z4} 
\beq
\frac{d P}{d
x}=
\int\limits_{0}^{\infty}\! d z\,
n(z)
\frac{d
\sigma_{eff}^{BH}(x,z)}{dx}\,,
\label{eq:3}
\eeq
where $x$ is the gluon fractional momentum, $n$ is the number density of 
the medium, and
\beq
\frac{d
\sigma_{eff}^{BH}(x,z)}{dx}=\mbox{Re}
\int d\r\,
\Psi^{*}(\r,x){\sigma}_{3}(\rho,x)\Psi_{m}(\r,x,z)\,.
\label{eq:4}
\eeq
Here ${\sigma}_{3}$ is the cross section for interaction of the 
$\bar{q}gq$ system with a scattering center.
The relative transverse separations in the $\bar{q}gq$ system are
$
\r_{g\bar{q}}=(1-x)\r$, $\r_{q\bar{q}}=-x\r \,$.
$\Psi(\r,x)$ is the 
light-cone wave function for the $q\rightarrow gq$ transition in vacuum, 
and $\Psi_{m}(\r,x,z)$ is the quark light-cone wave function in the medium at 
the longitudinal coordinate $z$ (we omit spin and color indices). 
The wave functions (modulo a color factor) read
\bea
\Psi(\r,x)&=& 
P(x)\left (\frac{\partial}{\partial \rho_{x}^{'}}
-i s_{g}\frac{\partial}{\partial \rho_{y}^{'}}\right )
\int\limits_{0}^{\infty}d\xi
\exp\left(-\frac{i\xi}{L_{f}}\right)
{\cal{K}}_{0}(\r,\xi|\r^{'},0)\Biggl |_{\r^{'}=0}\,,
\label{eq:5}\\
\Psi_{m}(\r,x,z)&=&
P(x)\left (\frac{\partial}{\partial \rho_{x}^{'}}
-i s_{g}\frac{\partial}{\partial \rho_{y}^{'}}\right )
\int\limits_{0}^{z}d\xi
\exp\left(-\frac{i\xi}{L_{f}}\right)
{\cal{K}}(\r,z|\r^{'},z-\xi)\Biggl |_{\r^{'}=0}\,,
\label{eq:6}
\eea
where $P(x)=
i\sqrt{{\alpha_{s}}/{2x}}
[s_{g}(2-x)+2s_{q}x]/2M(x)$, $s_{q,g}$ denote parton helicities,
${\cal{K}}$ is the Green's function
for the two-dimensional Hamiltonian
\beq
\hat{H}(z)=-\frac{1}{2M(x)}
\left(\frac{\partial}{\partial \r}\right)^{2}
          -i\frac{n(z){\sigma}_{3}(\rho,x)}{2}\,,
\label{eq:7}
\eeq
and 
\beq
{\cal{K}}_{0}(\r_{2},z_{2}|\r_{1},z_{1})=\frac{M(x)}{2\pi i(z_{2}-z_{1})}
\exp\left[\frac{iM(x)(\r_{2}-\r_{1})^{2}}{2(z_{2}-z_{1})}\right]
\label{eq:8}
\eeq
is the Green's function
for the Hamiltonian (\ref{eq:7}) with $v(\r,z)=0$, $M(x)=Ex(1-x)$,
and 
$
L_{f}={2Ex(1-x)}/{\epsilon^{2}}
$
with 
$
\epsilon^{2}=m_{g}^{2}(1-x)+m_{q}^{2}x
$. The gluon mass $m_{g}$ plays the role of infrared cutoff removing the
contribution from long wave gluons which cannot propagate in the QGP. It is
natural to take $m_{g}\sim \mu$. However, for large suppression 
which occurs at $E\rightarrow \infty$ the parton masses can simply be 
neglected. 

The three-body cross section can be written as \cite{NZ}
\beq
\sigma_{3}(\rho,x)=\frac{C_{A}}{2C_{F}}[\sigma_{2}((1-x)\rho)
+\sigma_{2}(\rho)-\frac{1}{N_{c}^{2}}\sigma_{2}(x\rho)]\,,
\label{eq:9}
\eeq
where $C_{A}=N_{c}$ is the octet color Casimir, $\sigma_{2}(\rho)$ is 
the dipole cross section for scattering of a $\bar{q}q$ pair 
on a color center. For the parametrization 
$\sigma_{2}(\rho)=C_{2}(\rho)\rho^{2}$ the factor $C_{2}$ is
\beq
C_{2}(\rho)=\frac{C_{T}C_{F}\alpha_{s}^{2}}{\rho^{2}}\int d\qb
\frac{[1-\exp(i\qb\r)]}{(q^{2}+\mu^{2})^{2}}\,\,.
\label{eq:10}
\eeq
Here $C_{T}$ is the color Casimir of the scattering center.
In the region $\rho\ll 1/\mu $ which dominates  the spectrum 
for strong suppression (\ref{eq:10}) takes the form
\beq
C_{2}(\rho)\approx\frac{C_{F}C_{T}\alpha_{s}^{2}\pi}{2}
\log\left(\frac{1}{\rho\mu}\right)\,.
\label{eq:11}
\eeq

The LLA consists in replacing $C_{2}(\rho)$ by $C_{2}(\rho_{eff})$,
where $\rho_{eff}$ is the typical value of $\rho$. 
This seems to be a reasonable procedure since 
$C_{2}(\rho)$ has only a slow logarithmic dependence on $\rho$.
Then $\sigma_{3}(\rho,x)=C_{3}(x)\rho^{2}$,
where $C_{3}(x)=C_{2}(\rho_{eff})A(x)$ with 
$A(x)=[1+(1-x)^{2}-x^{2}/N_{c}^{2}]C_{A}/2C_{F}$, and
the Hamiltonian (\ref{eq:7}) takes the oscillator form with the frequency
$\Omega(x)=\sqrt{-iC_{3}(x)n/M(x)}$.
The value of $\rho_{eff}$ is connected with the gluon formation 
length, $l_{f}$, by the Schr\"odinger diffusion relation 
$\rho_{eff}^{2}\sim l_{f}/2M$.
$l_{f}$ is simply the typical scale of
$\xi$ in (5), (\ref{eq:6}) when the wave functions are 
substituted in (\ref{eq:4}).

Let us discuss the gluon emission at qualitative level. 
We begin by estimating $\rho_{eff}$ and $l_{f}$.
Let us first estimate these quantities for gluon emission
from a quark in an infinite medium. We will denote them 
as $\bar{\rho}_{eff}$ and $\bar{l}_{f}$.
They should also be related by the Schr\"odinger diffusion relation.
On the other hand, the absorption effects for 
the $\bar{q}gq$ system should become strong at the scale $\bar{l}_{f}$. 
It means that
$
\bar{l}_{f}nC_{3}\bar{\rho}_{eff}^{2}/2\sim 1\,.$
From these conditions one gets
$
\bar{\rho}_{eff}\sim [E_{a}x(1-x)nC_{3}]^{-1/4}$
and 
$
\bar{l}_{eff}\sim 2\sqrt{{E_{a}x(1-x)}/{nC_{3}}}
$.
These estimates are valid when 
$\bar{\rho}_{eff}\lsim 1/\epsilon$ and $\bar{l}_{f}\lsim L_{f}$.
  
Now we turn to the gluon emission from a quark produced inside a finite-size 
medium.
In this case in the high-energy limit qualitatively 
two different situations are possible. 
The first regime gets for the gluons with $x$ such 
that $\bar{l}_{f}\lsim L$. In this case the finite-size effects play a 
marginal role, and $\rho_{eff}\sim \bar{\rho}_{eff}$. The spectrum can 
roughly be calculated using the 
effective Bethe-Heitler cross section for the infinite medium. 
We call this regime the infinite medium regime.
The second regime occurs for the gluons for which $\bar{l}_{f}\gsim L$.
In this case $\rho_{eff}\sim \rho_{d}(L)$, where 
$\rho_{d}(L)=\sqrt{L/2M}$ is simply the diffusion
radius on the scale of the quark path length inside the medium.
In this regime the effective Bethe-Heitler cross section
is chiefly controlled by the finite-size effects. We will call this regime
the diffusion regime.
Thus we can write for the above two regimes 
\beq
\rho_{eff}\sim \mbox{min}(\bar{\rho}_{eff},\rho_{d}(L),1/\epsilon)\,.
\label{eq:12}
\eeq
Here we have taken into account that $\rho_{eff}\lsim 1/\epsilon$.
In terms of $x$ the infinite medium regime
occurs at $x\lsim \delta$ and $(1-x)\lsim \delta$, and the diffusion regime
gets at $\delta \lsim x\lsim (1-\delta)$, where
\beq
\delta\sim \frac{nC_{3}L^{2}}{4E}\,.
\label{eq:13}
\eeq

For the sake of definiteness, below we discuss only the region $x\lsim 0.5$.
At $x\gsim\delta$ the probability of interaction of the $\bar{q}gq$ 
system with the medium (it is of the order of $n\sigma_{3}(\rho_{d},x)L$)
becomes small. Thus, it is clear that in the developed diffusion regime
the spectrum is dominated by the $N=1$ scattering.
It is surprising that this turns out to 
be in apparent contradiction with prediction of the LLA. 
The LLA spectrum can be obtained  
using in (\ref{eq:6}) the oscillator Green's function. For zero 
parton masses it gives
\beq
\frac{d P}{dx}=-\frac{2G(x)}{\pi}\mbox{Re}\int\limits_{0}^{L}dz
\int\limits_{0}^{z}d\xi\frac{\Omega^{2}}{\cos^{2} \Omega\xi}=
\frac{2G(x)}{\pi}\ln|\cos\Omega L|\,,
\label{eq:14}
\eeq
where
$
G(x)
=
{\alpha_{s}C_{F}}[1-x+x^{2}/{2}]/x\,$.
This spectrum has been derived in \cite{BDMS}.
Note that $|\Omega L|\sim 1$ at $x\sim \delta$.
For the diffusion regime from (\ref{eq:14}) one gets
\beq
\frac{d P}{dx}\Biggl |_{x\gg\delta}^{LLA}\approx 
\frac{G(x)C_{3}^{2}n^{2}L^{4}}{8\pi E^{2}x^{2}(1-x)^{2}}\,.
\label{eq:15}
\eeq
Since the right-hand side of (\ref{eq:15}) $\propto n^{2}$ it is clear that
it corresponds to the $N=2$ term.
Thus one sees that the $N=1$ contribution is simply absent in the LLA.

The fact that the LLA fails in the diffusion regime can be directly 
seen from calculation of the $N=1$ contribution.  
To obtain it one should 
use in (\ref{eq:6}) the free Green's function (\ref{eq:8}). 
Then in the massless limit (\ref{eq:4}) gives
\beq
\frac{d
\sigma_{eff}^{BH}(x,z)}{dx}\Biggl |_{N=1}
=\frac{G(x)A(x)M(x)}{2\pi}\mbox{Im}
\int\limits_{0}^{z}\frac{d\xi}{\xi^{2}}\int\limits_{0}^{\infty} d\rho^{2}
\rho^{2}C_{2}(\rho)\exp\left(\frac{iM(x)\rho^{2}}{2\xi}\right)\,.
\label{eq:16}
\eeq
For $C_{2}(\rho)=\mbox{const}$ the $\rho^{2}$-integral 
in (\ref{eq:16}) has zero imaginary part, and the right-hand side of 
(\ref{eq:16}) is also zero.
On the other hand, using (\ref{eq:11}) one gets from (\ref{eq:16})
\beq
\frac{d\sigma_{eff}^{BH}(x,z)}{dx}\Biggl |_{N=1}=
\frac{\alpha_{s}^{2}\pi C_{T}C_{F}G(x)A(x)z}{4E x(1-x)}\,.
\label{eq:17}
\eeq
Then (\ref{eq:3}) yields
\beq
\frac{d P}{dx}\Biggl |_{N=1}=
\frac{\alpha_{s}^{2}\pi C_{T}C_{F}
G(x)A(x) n L^{2}}{8E x(1-x)}\,.
\label{eq:18}
\eeq

Let us see why the LLA fails in momentum representation in which 
(\ref{eq:4}) reads
\beq
\frac{d\sigma_{eff}^{BH}(x,z)}{dx}=\frac{\alpha_{s}^{2}C_{T}C_{F}A(x)}
{(2\pi)^{2}}\mbox{Re}
\int d\pb d\qb \frac{[\Psi^{*}(\pb,x)-\Psi^{*}(\pb-\qb,x)]\Psi_{m}(\pb,x,z)}
{(q^{2}+\mu^{2})^{2}}\,.
\label{eq:19}
\eeq
In the massless limit from (\ref{eq:19}) one can obtain
\beq
\frac{d\sigma_{eff}^{BH}(x,z)}{dx}\Biggl |_{N=1}
=\frac{\alpha_{s}^{2}C_{T}C_{F}G(x)A(x)}
{2\pi}
\int d p^{2} d q^{2}  \frac{F(p,q) }
{(q^{2}+\mu^{2})^{2}}\,,
\label{eq:20}
\eeq
\beq
F(p,q)=
\mbox{Re}\,
\frac{1}
{p^{2}}\cdot
\left[1-\exp\left(-\frac{iz p^{2}}{2M(x)}\right)\right]
\cdot\int\limits_{0}^{2\pi}d\phi
\frac{\qb(\qb-\pb)}{(\qb-\pb)^{2}}
\,,
\label{eq:21}
\eeq
where $\phi$ is the angle between $\qb$ and $\pb$.  The logarithmic situation  
with dominance of $q^{2}\ll p^{2}$ would correspond 
to $F(p,q)\propto q^{2}$ at $q^{2}\ll p^{2}$. 
However, the azimuthal $\phi$ integral in (\ref{eq:21}) equals 
$2\pi\theta(q^{2}-p^{2})$, and the process is dominated by hard $t$-channel 
exchanges with $q^{2}>p^{2}\sim 2M(x)/z$.
After integrating over  $p^{2}$ and $q^{2}$ 
in (\ref{eq:20}) one reproduces (\ref{eq:17}). 

It must be emphasized that the LLA fails only in the diffusion regime.
But it is a good approximation in the infinite medium regime when
$\Psi_{m}$ falls off rapidly at the scale much smaller than $\rho_{d}(L)$.
It is also worth noting that the boundary (\ref{eq:13}) beyond which the 
diffusion regime occurs is obtained assuming that in the infinite medium
regime LPM suppression is strong 
(it means that $\bar{\rho}_{eff}(x\sim \delta)\ll 1/\epsilon$). 
It is possible that there the Bethe-Heitler
situation takes place. One can easily show that in this case 
$\delta\sim L\mu^{2}/2E$. Thus, in general, the diffusion regime
occurs for the gluons with energy $\omega\gsim \omega_{cr}$, where
\beq
\omega_{cr}\sim \mbox{max}\left(\frac{nC_{3}L^{2}}{4},
\frac{L\mu^{2}}{2}\right)\,.
\label{eq:22}
\eeq

Let us now discuss the energy loss. It can be written as
\beq
\Delta E=\int\limits_{\mu}^{\omega_{cr}}d\omega \omega\frac{dP}{d\omega}
+\int\limits_{\omega_{cr}}^{\omega_{max}}d\omega \omega\frac{dP}{d\omega}
\,.
\label{eq:23}
\eeq
One can show that the first term in (\ref{eq:23}) does not depend on energy, 
and is of the order of $\Delta E_{BDMS}$ (\ref{eq:1}) 
for both the LPM and Bethe-Heitler situations. 
At $E\rightarrow \infty$ the energy loss is dominated by the 
second term in (\ref{eq:23}) 
which grows logarithmically with $E$. Then, using (\ref{eq:18})
to the logarithmic accuracy one can obtain in the high-energy limit
\beq
\Delta E=\frac{C_{F}\alpha_{s}}{4}\frac{L^{2}\mu^{2}}{\lambda_{g}}
\log\frac{E}{\omega_{cr}}\,.
\label{eq:24}
\eeq
Here we have used 
$\alpha_{s}^{2}\pi C_{F}C_{T}A(0)n/2\mu^{2}=1/\lambda_{g}$.
Note that since $L\gg 1/\mu$ from (\ref{eq:22}) it follows that 
always $\omega_{cr}\gg \mu$. The qualitative estimates 
(including the region $\omega\lsim \omega_{cr}$) show that the 
appearance of $\omega_{cr}$ in the logarithm in (\ref{eq:24}) instead of 
$\mu$ in (\ref{eq:2}) for RHIC conditions ($L\sim 4$ fm) 
can suppresses the energy loss at $E\sim 10$ GeV by a factor of 
$\sim 0.5$. For SPS conditions ($L\sim 2$ fm) the suppression is not strong
($\sim 0.7-0.8$ at $E\sim 5$ GeV).
The above estimates are obtained for the plasma temperature $T=250$ MeV.
Note that the absence of $\omega_{cr}$ in the GLV prediction (\ref{eq:2})
is connected with the neglect in \cite{GLV} of the mass 
effects in evaluating the phase factor which controls the 
interference for gluon emission from different points of the quark trajectory.

The above analysis is valid for the gluon emission from a fast
gluon as well. In this case in (\ref{eq:24}) $C_{F}$ should be 
replaced by $2C_{A}$ (here the factor $2$ comes from 
symmetry of the spectrum with respect to change 
$x\leftrightarrow(1\!-\!x)$).

I am grateful to J.~Speth for the hospitality
at FZJ, J\"ulich, where this work was completed.   
This work was partially supported by the grants INTAS
97-30494 and DFG 436RUS17/45/00.


\end{document}